\documentclass{ifacconf}
\makeatletter
\let\old@ssect\@ssect 
\makeatother

\usepackage{natbib}
\usepackage{hyperref}

\makeatletter
\def\@ssect#1#2#3#4#5#6{%
  \NR@gettitle{#6}
  \old@ssect{#1}{#2}{#3}{#4}{#5}{#6}
}
\makeatother

\usepackage{graphicx}
\usepackage{amsmath}
\usepackage{amssymb}
\usepackage{amsfonts}
\usepackage{algorithmic}
\usepackage{graphicx}
\usepackage{textcomp}

\usepackage{epstopdf}
\usepackage{psfrag}    
\usepackage{tikz}
\usepackage[nolist]{acronym}
\usepackage{pgfplots}
\usepackage{diffcoeff}  

\pgfplotsset{compat=1.17}
\usepgfplotslibrary{statistics}

\acrodefplural{NN}[NN's]{\emph{Neural Networks}}
\acrodefplural{SDN}[SDN's]{\emph{State Derivative Normalizations}}

\begin{acronym}
    \acro{RNN}{\emph{Recurrent Neural Networks}}
    \acro{LSTM}{\emph{Long-Short Term Memory Networks}}
    \acro{NODE}{\emph{Neural Ordinary Differential Equations}}
    \acro{ODE}{\emph{Ordinary Differential Equations}}
    \acro{RKNN}{\emph{Runge-Kutta Neural Networks}}
    \acro{LTC}{\emph{Liquid Time Constant Neural Networks}}
    \acro{NN}{\emph{Neural Network}}
    \acro{ReLU}{\emph{Rectified Linear Function}}
    \acro{DAE}{\emph{Differential Algebraic Equations}}
    \acro{MSE}{\emph{Mean Squared Error}}
    \acro{TPE}{\emph{Tree-Structured Parzen Estimator}}
    \acro{TSEM}{\emph{Truncated Simulation Error Minimization}}
    \acro{BLA}{\emph{Best Linear Approximation}}
    \acro{RMSE}{\emph{Root Mean Squared Error}}
    \acro{CTS}{\emph{Cascaded Tank System}}
    \acro{BLA}{\emph{Best Linear Approximation}}
    \acro{NLSS}{\emph{Nonlinear State Space Model}}
    \acro{CT}{\emph{Continuous-Time}}
    \acro{DT}{\emph{Discrete-Time}}
    \acro{DFT}{\emph{Discrete Fourier Transform}}
    \acro{SDN}{\emph{State Derivative Normalization}}
    \acro{PEM}{\emph{Prediction Error Minimization}}
\end{acronym}

\begin{document}
\begin{frontmatter}

\title{State Derivative Normalization for Continuous-Time Deep Neural Networks}

\vspace{-5pt}

\thanks[footnoteinfo]{The project RACKET supported this research under grant 01IW20009 by the German Federal Ministry of Education and Research.}

\author[First]{Jonas Weigand} 
\author[Second]{Gerben I. Beintema} 
\author[Third]{Jonas Ulmen}
\author[Third]{Daniel Görges}
\author[Second]{Roland Tóth}
\author[Second]{Maarten Schoukens}
\author[First]{Martin Ruskowski}

\vspace{-5pt}

\address[First]{\footnotesize{Chair of Machine Tools and Control Systems, RPTU Kaiserslautern, and the German Research Center for Artificial Intelligence, Kaiserslautern, Germany (e-mail: jonas@weigand-weigand.de and martin.ruskowski@rptu.de, ORCID: 0000-0001-5835-3106, 0000-0002-6534-9057)}}

\address[Second]{\footnotesize{ Control Systems (CS) Group at the Department of Electrical Engineering, Eindhoven University of Technology, Netherlands. R. Tóth is also affiliated to the Systems and Control Laboratory at the Institute for Computer Science and Control, Budapest, Hungary (e-mail: g.i.beintema@tue.nl, r.toth@tue.nl and m.schoukens@tue.nl, ORCID: 0000-0002-7822-6283, 0000-0001-7570-6129, 0000-0002-4904-1255)}}

\address[Third]{\footnotesize{ Institute for Electromobility, RPTU Kaiserslautern, Germany (e-mail: jonas.ulmen@rptu.de and daniel.goerges@rptu.de, ORCID: 0000-0003-1597-1523, 0000-0001-5504-0972)}}

This work has been accepted for presentation at the $20^{th}$ IFAC Symposium on System Identification 2024.

\begin{abstract}
The importance of proper data normalization for deep neural networks is well known. However, in continuous-time state-space model estimation, it has been observed that improper normalization of either the hidden state or hidden state derivative of the model estimate, or even of the time interval can lead to numerical and optimization challenges with deep learning based methods. This results in a reduced model quality. In this contribution, we show that these three normalization tasks are inherently coupled. Due to the existence of this coupling, we propose a solution to all three normalization challenges by introducing a normalization constant at the state derivative level. We show that the appropriate choice of the normalization constant is related to the dynamics of the to-be-identified system and we derive multiple methods of obtaining an effective normalization constant. We compare and discuss all the normalization strategies on a benchmark problem based on experimental data from a cascaded tanks system and compare our results with other methods of the identification literature.
\end{abstract}

\begin{keyword}
System Identification, Continuous-Time, Neural Ordinary Differential Equations, Nonlinear State-Space
\end{keyword}

\end{frontmatter}

\maketitle

\vspace{-10pt}

\section{Introduction}
\label{sec:intro}

Machine learning is a powerful tool for time series modeling and system identification \citep{Schoukens.2019}. Models can be categorized into \ac{DT} models and \ac{CT} models such as \ac{NODE} \citep{Chen.2018}, \ac{RKNN} \citep{Wang.1998}, Deep Encoder Networks \citep{Beintema.2023} or \ac{LTC} \citep{Hasani.2020}. In control engineering, \ac{CT} models are often preferred, as they are closely related to physical models, they are often found more attractive for designing controllers since shaping performance with such models is more intuitive, and they can be used under irregular sampling. 

Data normalization is a key data preprocessing step in modern machine learning approaches. The core idea of normalization is to map the input and output data to a numerically favorable range (e.g. zero-mean and standard deviation of 1)~\citep{Ba.2016}. This is important to (\emph{i}) comply with the underlying assumption in many parameter initialization strategies for \acp{NN} (e.g., Xavier initialization \citep{Glorot.2010}) and to (\emph{ii}) numerically improve the effective numerical range of the gradients for the nonlinear activation functions~\citep{ioffe2015batch}.

Prior research demonstrates the significant enhancement in model performance through \ac{SDN} as evidenced by \citep{Weigand.2021b, Beintema.2023}. Despite these advancements, a comprehensive exploration and analysis of \acp{SDN} mechanisms remain absent from published literature. In addition, practical methodologies to estimate the normalization factor are missing in the literature. 



\begin{itemize}
    \item Consequently, this paper analyses the state, the state derivative, and the time normalization as well as their coupling. Additional graphical interpretation and the connection to \ac{ODE} solvers are presented. 

    \item Furthermore, three approaches are proposed to estimate this normalization factor and these approaches are compared on a well-studied benchmark example.
\end{itemize}

The remainder of the paper is structured as follows. Section~\ref{sec:identification} introduces the considered identification problem. We present multiple interpretations of the use of a normalization constant in Section~\ref{sec:interpretation}. Section~\ref{sec:methods} defines the criteria for an effective normalization and provides three methods for practical implementation. Section~\ref{sec:results} applies the method to a real-world benchmark problem, and in Section~\ref{sec:con} concluding remarks on the established results are made.

\section{The Identification Problem}
\label{sec:identification}
\subsection{Data-generating System}
Consider a 
nonlinear system that is represented by a nonlinear state-space representation
\begin{subequations}\label{eq:system-equations}
    \begin{align}
        \dot{x}(t) & = f\left(x(t), u(t)\right)  \label{eq:system_state} \\ 
        y_k &= h(x_k, u_k) + w_k,  \\
        x(0) &= x_0,
    \end{align}
    \label{eq:dx_dt}
\end{subequations}
\noindent where $t \in \mathbb{R}_{\ge 0}$ is the continuous time, $x(t) \in \mathbb{R}^{n_\mathrm{x}}$ is the state associated with \eqref{eq:dx_dt} with $n_\mathrm{x} \in \mathbb{N}$, $x_0\in \mathbb{R}^{n_\mathrm{x}}$ being the initial state, $u(t) \in \mathbb{R}^{n_\mathrm{u}}$ is the exogenous input, and $f:\mathbb{R}^{n_\mathrm{x}\times n_\mathrm{u}}\rightarrow \mathbb{R}^{n_\mathrm{x}}$ is locally Lipschitz continuous ensuring that all solutions of $x$ are forward complete. With respect to the output equation, ${y}(k) \in \mathbb{R}^{n_\mathrm{y}}$ stands for the sampled system output at time moments $k T_\mathrm{s}$ with $T_\mathrm{s}>0$ being the sampling interval and $k\in\mathbb{Z}_{\ge 0}$,  $x_{k} = x(k T_\mathrm{s})$ and $u_{k} = u(k T_\mathrm{s})$, while $w_k \in \mathbb{R}^{n_\mathrm{y}}$ is an i.i.d. zero-mean white noise process with finite variance $\Sigma_w$. Such a hybrid process and noise model is often used in \ac{CT} nonlinear system identification \citep{Schoukens.2019}.


\subsection{Identification}

For the considered system class, the \ac{CT} model identification problem can be expressed for a given data set of measurements
\begin{equation}
    D_N = \{(u_0, y_0), (u_1, y_1), ... ,(u_{N-1}, y_{N-1})\},
\end{equation}
generated by \eqref{eq:dx_dt}, with unknown $w_k$, $x(t)$, $\dot{x}(t)$, and initial state $x_0$, as the following optimization problem (a.k.a. $\ell_2$ simulation loss minimization)
\begin{equation}
\label{eq:full-sim-optim}
\begin{aligned}
    \min_{\theta,\hat{x}_0} \quad & \frac{1}{N} \sum_{k=0}^{N-1} \| y_k - \hat{y}_k \|_2^2, \\
    \textrm{s.t.} \quad & \hat{y}_k = h_\theta(\hat{x}(kT_\mathrm{s})),\\
      & \dot{\hat{x}}(t) = f_\theta(\hat{x}(t),u(t)),   \\
\end{aligned}
\end{equation}
\noindent where $\hat{x}(t) \in \mathbb{R}^{n_\mathrm{x}}$ is the model state, $h_\theta$ and $f_\theta$ are the output and state-derivative functions parameterized by $\theta \in\mathbb{R}^{n_\theta}$. These two functions are represented by multi-layer feedforward \acp{NN} within this paper. The state network $f_\theta(\cdot)$ input layer nodes are considered as model state $\hat{x}(t)$ and exogenous variables $u(t)$, and the output layer nodes are considered as model state derivative $\dot{\hat{x}}(t)$ \citep{schoukens2021improved, suykens1995nonlinear}. However, the presented results also hold when other function approximators such as polynomials are considered.


\section{Interpretation of the Normalization}
\label{sec:interpretation}

\textbf{\acf{SDN}} The idea is the introduction of a normalization factor $\tau$, which scales the state-derivative equation of the state-space model during training:

\begin{equation}
    \dot{\hat{x}}(t) = \frac{1}{\tau} f_{_\mathrm{NN}}\!\left(\hat{x}(t), u(t)\right)
    \label{eq:scaled_network_2}
\end{equation}


The normalization can be implemented as a scalar normalization $\tau \in \mathbb{R}_{>0}$, or as a vector $\tau \in \mathbb{R}^{n_\mathrm{x}}_{>0}$ corresponding to elements of the output of the hidden state network $f_{_\mathrm{NN}}$. In the latter case, \eqref{eq:scaled_network_2} is defined as $\dot{\hat{x}}(t) = \mathrm{diag}(\frac{1}{\tau_1}, \ldots, \frac{1}{\tau_{n_\mathrm{x}}}) f_\theta(\hat{x}(t),u(t))$. This linear scaling can also be interpreted as changing the weight initialization of the output layer of $f_{_\mathrm{NN}}(\cdot, \cdot)$. 


It is clear that $\tau$ in \eqref{eq:scaled_network_2} scales the hidden state derivative. Hence, it can be used to normalize the state derivatives. Next, we show that the derivative normalization is inherently coupled to the scaling of the state $x$ and the scaling of the time $t$.

\textbf{State Normalization.} It is also possible to rewrite \eqref{eq:scaled_network_2} as:
\begin{subequations}
    \begin{align}
        {\frac{d(\tau \hat{x}(t))}{dt}}    & = {f_{_\mathrm{NN}}}(\hat{x}(t), u(t))                \\
        \tilde{x}(t)                 & \triangleq \tau \hat{x}(t)                             \\
        {\frac{d\tilde{x}(t)}{dt}} & = {f_{_\mathrm{NN}}}(\tilde{x}(t)/ \tau, u(t))
    \end{align}
    \label{eq:state_norm}
\end{subequations}
In this way, normalization can be interpreted as the normalization of the magnitude of the state.

\textbf{Time-Based Normalization.} We can rewrite \eqref{eq:scaled_network_2} as
\begin{subequations}
    \begin{align}
        {\frac{d\hat{x}(t)}{d(t / \tau)}}             & = {f_{_\mathrm{NN}}}(\hat{x}(t), u(t)),                            \\
        \tilde{t}                               & \triangleq t / \tau ,                                          \\
        {\frac{d\hat{x}(\tilde{t} \tau)}{d\tilde{t}}} & = {f_{_\mathrm{NN}}}(\hat{x}(\tilde{t} \tau), u(\tilde{t} \tau)),
    \end{align}
    \label{eq:time_norm}
\end{subequations}
\noindent which suggests that the normalization can be viewed as a scaling of time by a factor $\tau$.

This time rescaling can also be viewed as changing the effective integration length in \ac{ODE} solvers. For instance,
\begin{subequations}
\begin{align}
    \hat{x}(t) &= \hat{x}(0) + \int_0^t \frac{1}{\tau} f_{_\mathrm{NN}} ( \hat{x}(t'), u(t'))\ dt' \\
    &= \hat{x}(0) + \int_0^{t/\tau}\!\!\! f_{_\mathrm{NN}} ( \hat{x}(\tau \tilde{t}'), u(\tau \tilde{t}'))\ d\tilde{t}' 
\end{align}
\end{subequations}
using the substitution of $\tilde{t}' = t' / \tau $. This integration by substitution shows that the normalization can be viewed as rescaling the effective integration time from $t$ to $t/\tau$. This can be translated directly to a wide range of numerical integration schemes by rescaling the integration length and step size simultaneously. 

\begin{figure}[!htb]
    \centering
    \input{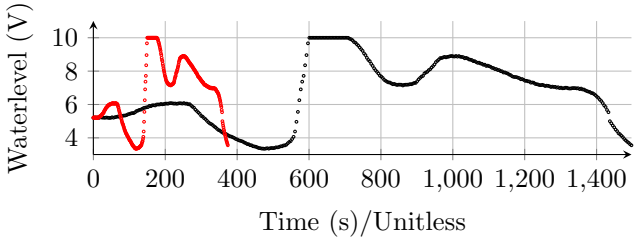}
    \vspace{-15pt}
    \caption{Time scaling of the output measurement of the \acf{CTS} benchmark \citep{Schoukens.2016b}. Black: Original measurement (unit seconds). Red: Time-domain scaled data (unitless).}
    \label{fig:time_domain_nor_conti}
\end{figure}

Graphical intuition about the time rescaling can be gained from output data of the \ac{CTS} benchmark in Fig.~\ref{fig:time_domain_nor_conti} \citep{Schoukens.2016b}. The measured data (black) of the water level is given at a sample rate of $T_\mathrm{s} = 4.0 \, \mathrm{s}$, the scaled data (red) is transferred to a sample time of $T_m = T_\mathrm{s}/\tau = 1.0$, while the number of time steps remains unchanged. The original data is transferred to the scaled time domain, where the model is trained and evaluated, and the results are transferred back to the original time grid.

It is known that the stiffness of an ODE system is subject to temporal variation. However, employing a constant (linear) normalization across all stiffness regions is essential from a practical model inference point of view. Moreover, averaging across multiple regions increases the robustness of the estimation by providing a broader range of suitable normalization factors. Our subsequent experiments demonstrate that there exists a broad valley of good normalization factors (Fig.~\ref{fig:box_result}). Remarkably, all normalization factors within this range lead to high-performance models.


The key takeaway of representations \eqref{eq:scaled_network_2}, \eqref{eq:time_norm} and \eqref{eq:state_norm} is that \ac{SDN} collectively influences the hidden state, the hidden state derivative, and the scaling of time $t$. 

\section{Estimation of the Normalization Factor}
\label{sec:methods}

Since the system dynamics are unknown a priori, selecting a normalization factor $\tau$ that results in a favorable numerical performance of the training algorithm is challenging a priori, it has to be estimated during training, from the available data, or by heuristics. We propose three methods:
\begin{itemize}
    \item make the normalization factor a trainable parameter,
    \item estimate $\tau$ using cross-validation,
    \item use a heuristic-based approach starting from a linear approximation of the system.
\end{itemize}

\textbf{Trainable Parameter.} The normalization factor $\tau$ can be added to the set of parameters present in the optimization problem \eqref{eq:full-sim-optim}. This results in a simultaneous optimization of $\tau$, $\theta$, and $\hat{x}_0$ under the $\ell_2$ simulation loss. For this method, a vectorized normalization leads to more optimization parameters than using a single scalar.

In terms of implementation, $\tau > 0$ should hold to fulfil its role as a magnitude normalization. As most \ac{NN} libraries do not provide constrained optimization one can implement
\begin{equation}
    \hat{\tau} = \mathrm{max}(\epsilon, \tau )
    \label{eq:positive_trainable_parameter}
\end{equation}
with a small positive number $\epsilon \in \mathbb{R}_{>0}$. Empirical evidence demonstrates that the trainable normalization quickly reaches a stationary value.



From a pure mathematical perspective, making $\tau$ a trainable parameter may seem to have little impact because the expressive power of the \ac{NN} could easily compensate for it. Practically, the optimizer would need to adapt a large quantity of parameters instead of a single normalization factor. Additionally, traditional \ac{NN} initialization methods assume inputs and outputs are within a specific range. We consolidated the normalization dependency into a single trainable parameter $\tau$, making the training process simpler and more robust, as demonstrated in our experiments.

Overall, this method offers high adaptability across different data sets and \ac{NN} configurations, an easy implementation, and demands only little additional computational load.


\textbf{Cross-validation.} The normalization factor $\tau$ can be considered as a hyperparameter and estimated using cross-validation methods. For example, \eqref{eq:full-sim-optim} is optimized using a training data set for a chosen set of values of $\tau$, a validation data set is used subsequently to determine the best performing $\tau$ value. This method is universally applicable to nonlinear systems, is easy to implement if a hyperparameter tuning is already available, and, given sufficient function evaluations, ensures global optimality.

\textbf{Heuristic approach based on an approximate model.} This heuristic is derived based on a guarantee of the existence of a normalized model formalized by \citep{Beintema.2023} which states;

\begin{thm}
    Given a input trajectory $u(t)$, a non-constant state trajectory $x(t)$ which satisfies $\dot x (t) = f(x(t),u(t))$ as in \eqref{eq:system-equations} for all $t \in [0,L]$ than there exists a $\tau \in \mathbb{R}^+$ and a scalar state transformation $\gamma \hat {x}(t) = x(t)$ such that both the model state trajectory $\hat{x}(t)$ and model derivative $f_{\textrm{NN}}$ given by $\dot{\hat{x}}(t) = \tfrac{1}{\tau} f_{\textrm{NN}}(\hat{x}(t),u(t))$ as in \eqref{eq:scaled_network_2} are normalized as 
    \begin{subequations}
        \begin{align}
            \mathrm{var}(\hat{x}) &\triangleq \frac{1}{L} \int_0^L \frac{1}{n_\mathrm{x}} \| \hat{x}(t)\|_2^2 dt = 1, \\
            \mathrm{var}(f_{\textrm{NN}}(\hat{x},u)) &= 1.
        \end{align}
    \end{subequations}
    
\end{thm}
\begin{pf}
    With 
    \begin{align}
        \gamma &= \sqrt{\mathrm{var}(x)} \\ \tau &= \sqrt{\mathrm{var}( x)/\mathrm{var}(\dot x)} \label{eq:variance2}
    \end{align}
    the normalization conditions are satisfied, as shown below
\begin{align*}
    \mathrm{var}(\hat{x}) &= \mathrm{var}(x/\gamma) = \mathrm{var}(x)/\gamma^2 = 1,\\
    \mathrm{var}(f_{\textrm{NN}}(\hat{x},u)) &= \mathrm{var}(\tau \dot{\hat{x}}) = \mathrm{var}(\tau \dot{x} /\gamma) = \tau^2/\gamma^2 \mathrm{var}(\dot{x})=1.
\end{align*}
\end{pf}

We base our heuristic on the relationship of \eqref{eq:variance2}. We make two alterations \textit{(i)} the integral over time can be approximated by a summation over the time samples to make this tractable and \textit{(ii)} in \eqref{eq:variance2} we use an approximate model instead of the system equations.

We have observed that using a linear approximate model based on the \ac{BLA} of a nonlinear system can get a sufficiently accurate estimate of the optimal $\tau$. The \ac{BLA} offers a linear approximation of a nonlinear system, best in the mean-squared-error sense, based on measured input-output data \citep{pintelon2012system}. Estimating a \ac{BLA} of a potentially nonlinear system is commonly done using \ac{PEM}, minimizing a least squares prediction error. Using an estimated \ac{BLA}, an estimate of the state and state-derivatives, $x_\mathrm{BLA}$ and $\dot{x}_\mathrm{BLA}$, can be obtained to compute $\tau$ through \eqref{eq:variance2}: 
\begin{equation}
    \tau_\mathrm{BLA} = \sqrt{\mathrm{var}( x_\mathrm{BLA} )/ \mathrm{var}( \dot{x}_\mathrm{BLA} )}.
    \label{eq:estimate_tau_bla}
\end{equation}
This method is especially valuable for dynamic systems that can be reasonably well represented by a linear model for the considered range of excitation. Also note that any linear transformation that can be present in the \ac{BLA} state and state derivative estimate does not affect the resulting $\tau$ estimate in \eqref{eq:estimate_tau_bla}. Additionally, besides normalization, the \ac{BLA} can be used as a good \ac{NN} weight initialization candidate \citep{Schoukens.10.04.2020}.

 Frequency-domain analysis of this approach offers additional insight into the optimal choice of the normalization factor. Consider that the input signal has a periodicity of $N$ samples, then one can decompose the input into its \ac{DFT} components as:
\begin{gather}
    u_k = \frac{1}{N}\sum_{m=0}^{N-1} U_m e^{j \frac{2\pi}{N} m k} \\
    U_m = \sum_{k=0}^{N-1} u_k  e^{-j \frac{2\pi}{N} m k }
\end{gather} 
Both the variance of the  state and state-derivative can be expressed using these \ac{DFT} components as:
\begin{subequations}\label{eq:varxfourier}
\begin{align}
    \text{var} (x) &= \frac{1}{L} \int_0^L \|x(t)\|_2^2 dt \sim \sum_{m=-\infty}^\infty \| X_m \|_2^2\\
    \text{var} (\dot{x}) &= \frac{1}{L} \int_0^L \|\dot{x}(t)\|_2^2 dt \sim \sum_{m=-\infty}^\infty \omega_m^2 \| X_m \|_2^2
\end{align}
\end{subequations}
where $\omega_m = 2\pi\frac{m}{NT_s}$.

Since we are considering the \ac{BLA}, the Fourier components of the state can be written in terms of the input-to-state frequency response function $X_m = G(j\omega_m) U_m$, assuming a single input system for simplicity. Therefore, substituting this and \eqref{eq:varxfourier} into \eqref{eq:estimate_tau_bla} results in
\begin{equation}
    \tau_\mathrm{BLA} = \sqrt{\frac{ \sum_{m=-\infty}^\infty U_m^H G^H\! ( j\omega_m ) G( j\omega_m ) U_m }{ \sum_{m=-\infty}^\infty \omega_m^2 U_m^H G^H\! ( j\omega_m ) G( j\omega_m )U_m }},
    \label{eq:tau_freq}
\end{equation}
where $\cdot^H$ denotes the Hermitian operation.
Hence, if the signal $u(t)$ only consists of a single sine wave (i.e. only one nonzero $U_m$) then $\tau_\mathrm{BLA} = 1/\omega_m$. Furthermore, if $u(t)$ or $G(j \omega_m)$ has a finite bandwidth bounded by $\Omega$ then $\tau_\mathrm{BLA} \geq 1/\Omega$.

\section{Experimental Results}
\label{sec:results}

\begin{figure}
    \centering
    \includegraphics[width=0.45\linewidth]{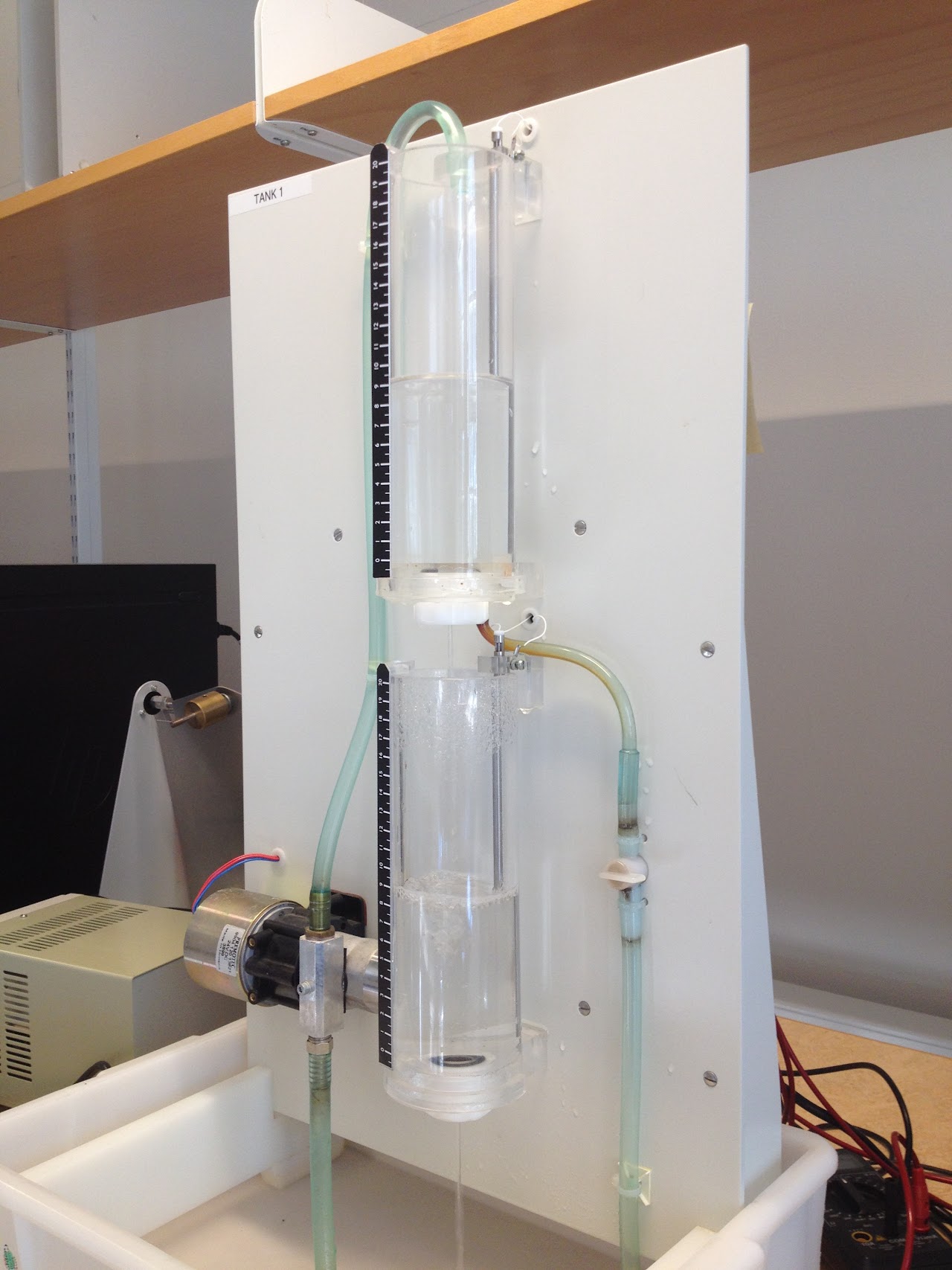}
    \caption{Picture of the \acf{CTS} \citep{Schoukens.2016b}.}
    \label{fig:cascaded_tanks}
\end{figure}

\textbf{Benchmark.} We apply the proposed methods to the \ac{CTS} benchmark \citep{Schoukens.2016b}. The setup is depicted in Fig.~\ref{fig:cascaded_tanks}. It consists of two vertically mounted tanks, where the upper one has a water inflow using a pump, and the water flows from the upper tank into the lower one. The task is to estimate a model that can predict the water level of the lower tank given a pump input sequence. The experiment incorporates an overflow of the tank, which introduces a hard saturation function.

\textbf{Model Configuration.} We chose the fixed-step Runge-Kutta 4 \ac{ODE} solver. The network ${f_{_\mathrm{NN}}}( \cdot, \cdot )$ is chosen  to consist of linear layers with matrices $A \in \mathbb{R}^{n_\mathrm{x} \times n_\mathrm{x}}$, $B \in \mathbb{R}^{n_\mathrm{x} \times n_\mathrm{u}}$ and two residual hidden layers with a Leaky ReLU activation function $\sigma( \cdot ) $ and $64$ hidden units for each layer. Weight matrices are denoted as $W_{(\cdot)}$ and bias terms as $b_{(\cdot)}$ with appropriate dimensions. Bias terms for the linear layer are disabled. An output network ${g_{_\mathrm{NN}}}( \cdot, \cdot )$ is also used with the same structure as the state network ${f_{_\mathrm{NN}}}( \cdot, \cdot )$. The overall \ac{NN} model can be written as:
\begin{subequations}
    \begin{align}
        {\frac{dx(t)}{dt}} &= {f_{_\mathrm{NN}}}( x(t), u(t) ) = A x(t) + B u(t) \\
        &+ W_{F1} \sigma \left( W_{F2} \sigma \left( W_{F3} \begin{bmatrix} x(t) \\ u(t) \end{bmatrix} + b_{F3} \right)  + b_{F2} \right)  \nonumber \\
        y(t) &= {g_{_\mathrm{NN}}}( x(t), u(t) ) = C x(t) + D u(t) \\
        &+ W_{G1} \sigma \left( W_{G2} \sigma \left( W_{G3} \begin{bmatrix} x(t) \\ u(t) \end{bmatrix} + b_{G3} \right)  + b_{G2} \right)  \nonumber
    \end{align}
    \label{eq:networks}
\end{subequations}
More details on the general model structure are given in \citep{schoukens2021improved, suykens1995nonlinear}. White box modeling of the \ac{CTS} would lead to $2$ states. Nevertheless, we set the number of states $n_\mathrm{x}=4$. It is observed that this results in better learning behavior. This is motivated by the effect of state augmentation \citep{Dupont.2019}, corresponding to the fact that increasing the state dimension allows to describe the system behavior with less complex nonlinearities involved (also expressed by the immersion concept in nonlinear system theory). The initial hidden states are obtained using a \emph{Deep Encoder Network} \citep{Beintema.2023} ${e_{_\mathrm{NN}}}( \cdot, \cdot )$ with $n_\mathrm{a} = n_\mathrm{b} = 5$. Weight matrices are initialized with a small random number chosen from a uniform distribution $\mathcal{U}[-0.01, 0.01]$, while bias terms are initialized to zero. The scalar normalization is initialized with $1$ and the vectorized normalization with $1/{n_\mathrm{x}}$. A barrier function $L_\mathrm{N}$ is applied to the linear layer $A$ of the state network to ensure negative definiteness \citep{Weigand.2021b}. This barrier minimizes drifts over long simulation horizons and is estimated using the differentiable Sylvester Criterion. Furthermore, we apply a \ac{DAE} network
\begin{align}
    &d_{_\mathrm{NN}}( x(t), u(t) ) =  \\
    &W_{D1} \sigma \left( W_{D2} \sigma \left( W_{D3} \begin{bmatrix} x(t) \\ u(t) \end{bmatrix} + b_{D3} \right)  + b_{D2} \right)  + b_{D1} \nonumber
\end{align}
to account for the hard state saturation, which is trained using an additional penalty function $L_\mathrm{D}$. It does not affect the forward model evaluation. The optimization problem is given by
\begin{equation}
    \begin{aligned}
        \min_{\theta} \quad & L_\mathrm{N} + L_\mathrm{W} + \\
                            &\sum_{k=\max(n_\mathrm{a},n_\mathrm{b})}^{K-J} \left( \sum_{j=0}^{J-1} \| y_{k+j} - \hat{y}_{k+j|k} \|_2^2  + L_{\mathrm{D}, j} \right), \nonumber \\
        \textrm{s.t.} \quad & \hat{y}_{k+j|k} = g_{_\mathrm{NN}}(x_{k+j|k}),                                                                               \\
                             x_{k+j+1|k} &= \text{ODE\_Solve}\left(x_{k+j|k}, u_{k+j}, \frac{1}{\tau} f_{_\mathrm{NN}}( \cdot, \cdot ), T_\mathrm{s} \right),                       \\
                            & x_{k|k} = e_{_\mathrm{NN}}(u_{k-1},...,u_{k-n_b}, y_{k-1}, ..., y_{k-n_a}),                                                  \\
                            & L_{D, j} = \lambda_\mathrm{D} \, \left(d_{_\mathrm{NN}}( x_{j|j}, u_{j|j})\right)^2,                                                \\
                            & L_\mathrm{N} = \begin{cases}
                                          0 & \textrm{if}\ A 	\prec 0 \\
                                          \lambda_\mathrm{N} & \textrm{otherwise}.
                                      \end{cases}, \\
                            & L_\mathrm{W} = \lambda_\mathrm{W} \| \theta \|_2^2
    \end{aligned}
    \label{eq:encoder-method}
\end{equation}

\noindent with $\lambda_\mathrm{N} = 10^{12}$ and $\lambda_\mathrm{D} = 10^3$. The bar notation $x_{k+j|k}$ reads as "The simulated state at $x_{k+j}$ starting at $k$ with initial state $x_{k|k}$". The notation $\| \cdot \|_2^2$ stands for the squared Euclidean norm. The sampling time is $T_\mathrm{s}=4$ s.

The implementation is written in Python, using PyTorch. Furthermore, the ADAM optimizer  is applied with unmodified configuration except for the learning rate, which is set to $0.003$ for the first $1{,}000$ steps, and to $ 0.0009$ until step $3{,}000$, and to $0.00027$ for all subsequent iterations. The maximum number of optimization steps is set to $20,000$. Regularization is obtained using weight decay of $\lambda_\mathrm{W} = 10^{-8}$ and early stopping when the best validation error does not improve for $2{,}000$ iterations. We do not apply \emph{Batch-Norm} or \emph{Dropout}. Input and output data are z-score normalized. As no explicit validation data set is given for \ac{CTS}, we apply the first $512$ time steps of the test data for early stopping. Test data is not accessed for any other reason. Training is performed in $64$ mini-batches with a sequence length of $J = 128$ steps using \ac{TSEM} \citep{Forgione.2021b}. 

\begin{figure}[htb]
    \centering
\begin{tikzpicture}
    \begin{axis}[
    boxplot,
    boxplot/draw direction=y,
    boxplot/whisker range={100},
    ymin=0.0,
    ymax=1.0,
    width=\linewidth,
    height=0.5\linewidth,
    yminorgrids,
    ymajorgrids,
    minor tick num=1,
    ylabel = Error $e_{RMSE}$ (V),
    xtick={1,3,5,6,7,8,9,10,11},
    xticklabels={
      0.0001,
      0.002,
      0.03,
      0.1,
      0.5,
      2.3,
      4.0,
      9.5,
      40},
      xlabel={Normalized time $T_s / \tau$ (unitless)},
    ]
    \addplot[black] table[x expr=\coordindex, y index=0]{
  1.5273049196577642
     0.6772278731175515
     0.6385900082843871
     0.6520362957995027
     1.224170585250497
     0.6529778290216233
     0.5774258700518742
     0.5854752110476938
     0.6240344866615348
     0.651380100597924
    0.6287038801269456
    0.653665432329054
    0.7797801576620277
     0.5991782108587521
     0.6719072588399754
     0.6241164626004828
     0.6504278502839318
     0.5875285363551589
     0.758726713579029
     0.6191481252618283
    };
    \addplot[black] table[x expr=\coordindex, y index=0]{
   0.611551625595797
     0.6818736611187678
     0.6219077351572899
     0.6141389423097634
     0.6520644519214361
     1.8991134417899205
     0.6435006611942153
     0.6264724815761277
     0.6102892522045976
     0.6533365039364561
   0.6631496596423584
     0.6120748261115656
     1.4716217336150514
     0.7205353662516616
     0.7242504692813955
     0.6464184771198587
     0.6325931796151474
     0.6476865535532389
     0.6101775856191002
     0.6679900828310562
    };
    \addplot[black] table[x expr=\coordindex, y index=0]{
   0.8295235119713347
     0.7385351496215382
     0.687519053915886
     0.7188951330370806
     0.5759494653237166
     0.66410669086663
     0.6848851667936846
     0.6939229167237349
     0.8028146377362533
     0.8891599874633619
   0.769333375253337
     0.7656909525547075
     0.6942987779193244
     0.6603523440378859
     0.7335536652534673
     0.6583686723475746
     0.6566326256119126
     0.593475080593854
     0.7280219559648169
     0.9795590635824716
    };
    \addplot[black] table[x expr=\coordindex, y index=0]{
   0.6075120635228819
     0.7126068606207353
     0.47361590417840554
     0.6873892542348116
     0.3835223516907977
     0.6130592363473932
     0.3647707933585184
     0.36307852976570953
     0.639574089321715
     0.4382091914331215
   0.6608077364857096
     0.7358111284400021
     0.47747401607133855
     0.6920188880274584
     0.35318297281678174
     0.637662411964048
     0.7402600276797728
     0.5642085492019917
     0.7301649463673984
     0.6784149795145573
    };
    \addplot[black] table[x expr=\coordindex, y index=0]{
   1.1700760840675417
     0.31480007232648094
     0.4174534257214567
     0.3404287092353927
     0.46362015232348736
     0.2990140733644252
     0.3488864510313661
     0.36457961847507586
     0.25660720276455834
     0.2236699738242023
   0.33010692508525374
     0.7791712485339005
     0.32672603153281765
     0.3467230432200227
     0.27693502064251374
     0.2990967720431053
     0.3527386621647674
     0.47296818575208915
     0.3204624574894095
     0.318215885504504
    };
    \addplot[black] table[x expr=\coordindex, y index=0]{
   0.2997702700717555
     0.39470625068113907
     0.30278604262955033
     0.41991107936287414
     0.2676158095453486
     0.22246657723282992
     0.28321571308377913
     0.2834146220366212
     0.42462159636958186
     0.352378861550091
   0.2909529425743262
     0.5917071649462359
     0.24489038435982216
     0.2929188164762765
     0.36819863230411737
     0.3155150500414327
     0.5288912375025998
     0.4848347354840026
     0.2588909502845991
     0.3214577793242123
    };
    \addplot[black] table[x expr=\coordindex, y index=0]{
   0.2803139791653193
     0.38136809536722893
     0.20530225327883864
     0.27542917085871677
     0.31882532047048434
     0.2824022290857308
     0.2913141838560751
     0.31746071175212137
     0.23377579403554768
     0.265709333894929
   0.22074482501797968
     0.32989449592064707
     0.2896904178475634
     0.20978775598780008
     0.24542796853210974
     0.3984876694534261
     0.2453103796488067
     0.29062419240937865
     0.2493183836780409
     0.22297616615734692
    };
    \addplot[black] table[x expr=\coordindex, y index=0]{
   0.25213300279654993
     0.32179976728857346
     0.23589560116417713
     0.37756481596597435
     0.2884356861068816
     0.3082700166710099
     0.2755384499992206
     0.3174959620956275
     0.25759610904769314
     0.30633990082234863
   0.2860795452607297
     0.26592771105452145
     0.2415299907720819
     0.27753441718766814
     0.2572825150721544
     0.26506229039031737
     0.31493194017821646
     0.2941311897311571
     0.2831148319241333
     0.2938830030193306
    };
    \addplot[black] table[x expr=\coordindex, y index=0]{
    2.2877216380341943
    1.4964212720560164
    0.6209970787230807
    2.721280767688099
    0.7717900040935154
    0.4097769610219908
    0.6866973460336846
    0.7251363443776532
    0.573843877512684
    15.198051870410136
    0.9546367840783639
    1.76688326792758
    0.625569758241321
    0.6262470213443653
    0.7331109460503301
    0.6216938306832366
    3.2866044070456586
    0.28283438480451195
    2.8938137713830563
    2.4075532151258643
    };
    \addplot[black] table[x expr=\coordindex, y index=0]{
   0.28095270993846533
     0.656757749997914
     0.6564192672436581
     0.2578644719382463
     0.9174360416079683
     0.3608719082166649
     0.2412227184881788
     0.9631087489345567
     0.3753292537038949
     0.4686046636485073
   0.642411079408436
     0.6584154895134833
     0.661605469981089
     0.7527087529696058
     0.40793796355814205
     0.5150509219058009
     0.7501112865613786
     1.6302505843799437
     0.5664769727489037
    100
    };
    \addplot[black] table[x expr=\coordindex, y index=0]{
   2.1486728932328747
     36.848564617830945
     2.1826925219196744
     2.1754907689313905
     1.2093201849605402
     2.122094621622658
     2.1748207065666874
     2.1470217845460597
     0.7662116055846072
    100
     2.1591899032810393
     2.0893323628140537
     0.5823682773964419
     2.090738965831058
     2.1221022066890485
     100
     100
     100
     100
     100
    };
    \end{axis}
\end{tikzpicture}
    \vspace{-10pt}
    \caption{$200$ independent experiments, each with the same configuration except for the random weight initialization and a fixed scalar normalization factor.
    We tested $10$ different normalization factors repeated $20$ times each. 
    The box plot displays the median, lower quartile, upper quartile, minimum and maximum values. Note that experiments with a normalization $T_\mathrm{s} / \tau = 40$ sometimes lead to unstable results, with \ac{RMSE} $ > 10^9$.}
    \label{fig:box_result}
\end{figure}
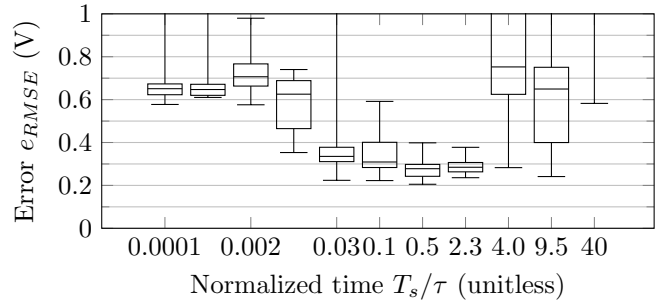

\begin{figure}[htb]
    \centering
    \input{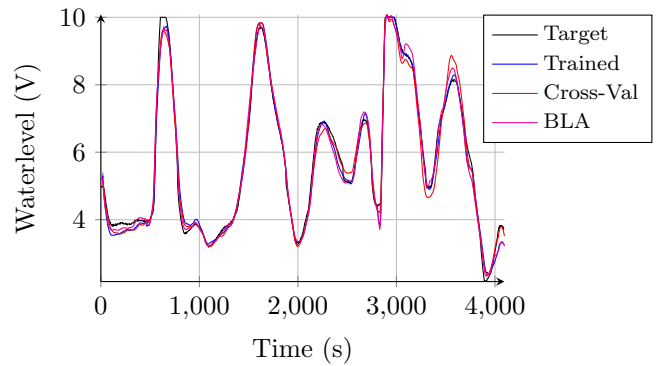}
    \vspace{-15pt}
    \caption{Simulation results on the \ac{CTS} benchmark \citep{Schoukens.2016b}. Results of the best models obtained with the trained normalization factor, the cross-validation method, and the \acl{BLA} are displayed.}
    \label{fig:sim_result}
\end{figure}

\textbf{Effect of Normalization.} We analyze the effect of the normalization factor $\tau$ given the \ac{CTS} benchmark, a fixed \ac{NN} configuration in simulation mode (free-run simulation/simulation error), and a fixed training pipeline throughout all experiments. In addition to the proposed state derivative normalization, the magnitude of the data (waterlevel) is normalized, too. Performance is measured in terms of the \ac{RMSE} of the simulation error w.r.t. the test data:
\begin{equation}
    e_\mathrm{RMSE} = \sqrt{ \frac{1}{K - \max(n_\mathrm{a},n_\mathrm{b})} \sum_{k=\max(n_\mathrm{a},n_\mathrm{b})}^{K-1} \|y_{k} - \hat{y}_k\|_2^2}. \nonumber
\end{equation}
Fig.~\ref{fig:box_result} displays a grid search for different scalar values of $T_\mathrm{s} / \tau$ fixed prior to the model training on a log scale ($T_\mathrm{s} = 4 \, \mathrm{s}$ for \ac{CTS}). Each experiment is repeated $20$ times to account for and observe the effect of randomness in the initial weights. We observe in Fig.~\ref{fig:box_result} that both the performance and variance of the results get worse for large and small values of $\tau$. Furthermore, Fig.~\ref{fig:box_result} indicates that there exists a desirable optimum. The optimum is different from the unscaled function estimation with $\tau=1$, which corresponds to $T_\mathrm{s} / \tau = 4.0$ for \ac{CTS} in Fig.~\ref{fig:box_result}. The median \ac{RMSE} for $T_\mathrm{s} / \tau = 4.0$ is $0.75 (V)$ and the mean \ac{RMSE} is $1.98 (V)$, considerably larger than the normalized result. The model outputs obtained with the proposed estimators for $\tau$ are displayed in Figure~\ref{fig:sim_result}.


 Using the cross-validation method to estimate the normalization, we observe in Fig.~\ref{fig:box_result} good results between $T_\mathrm{s} / \tau = 0.031$ and $T_\mathrm{s} / \tau = 2.276$, with a best \ac{RMSE} at $T_\mathrm{s} / \tau = 0.543$. For the trainable normalization method, we define a normalization vector $\tau$ as a trainable parameter and implement \eqref{eq:positive_trainable_parameter}. It is not optimized with hyperparameter tuning but jointly trained with all \ac{NN} weights, using the same learning rate. The normalization is initialized to $T_\mathrm{s} / \tau  = 0.1$. After training $20$ models, we obtain an average normalized time constant $T_\mathrm{s} / \tau  = 0.072 \pm 0.048 \, (\textrm{mean} \pm \textrm{std})$. Using the \ac{BLA} and \eqref{eq:estimate_tau_bla}, we estimate a normalization factor $T_\mathrm{s} / \tau = 0.054$, which matches the results in Fig.~\ref{fig:box_result}.


\textbf{Comparison to Literature.} Since several black-box identification methods have been applied to the \ac{CTS} in the literature, we can compare our method to the black-box approaches with the best performance (in terms of \ac{RMSE} on the test data). This comparison is provided in Table~\ref{tab:tank_res_sim}. The following approaches are compared: \citep{Relan.2017} estimates a \ac{BLA} and develops an unstructured \ac{NLSS} with different initialization schemes. A nonparametric Volterra series model is estimated in \citep{Birpoutsoukis.2018}. A nonlinear state-space model based on Gaussian processes is applied in \citep{Svensson.2017}. As a direct comparison, continuous-time \ac{NN} on this modeling problem has been applied in \citep{Beintema.2023, Forgione.2021b, Mavkov.2020, Weigand.2021b}. These works emphasize initial state estimation, fitting criteria, and model stability differently. It can be observed that the proposed approach outperforms all other black-box identification approaches.

\begin{table}[hb]
\begin{center}
\caption{Results for the Cascaded Tank Benchmark.}
\vspace{-5pt}
\label{tab:tank_res_sim}
    \begin{tabular}{ll}
                         & Test data      \\
        Method           & $e_{RMSE}$ (V) \\\hline
        Best Linear Approximation                       & 0.75           \\
        Truncated Volterra Model                        & 0.54           \\
        State-space with GP-inspired Prior              & 0.45           \\
        Integrated Neural Networks                      & 0.41           \\
        Soft-constrained Integration Method             & 0.40           \\
        Stable Runge-Kutta Neural Network               & 0.39           \\
        Nonlinear State Space Model                     & 0.34           \\
        Truncated Simulation Error Minimization         & 0.33           \\
        Deep Subspace Encoder                           & 0.22           \\\hline
        ours (SDN, Trained Parameter, best model)                                 & 0.2151    \\
        ours (SDN, Trained Parameter, mean $\pm$ std)                             & 0.2977 $\pm$  0.1259    \\
        ours (SDN, Cross-Validation, best model)                                  & 0.2054    \\
        ours (SDN, Cross-Validation, mean $\pm$ std)                              & 0.2777 $\pm$ 0.052    \\
        ours (SDN, \ac{BLA}, best model)                                          & 0.2253    \\
        ours (SDN, \ac{BLA}, mean $\pm$ std)                                      & 0.2633 $\pm$ 0.0284    \\\hline
    \end{tabular}
\end{center}
\end{table}

\vspace{-5pt}
\section{Conclusion}
\label{sec:con}
\vspace{-5pt}

We have shown the importance of proper state normalization when considering continuous-time modeling with state-space \acp{NN}. This is handled by introducing a normalization constant in front of the state derivative network. We have provided a state domain, a state derivative domain, and a time domain interpretation of this concept, and showed the beneficial effects of such a normalization. To estimate the appropriate normalization constant, three approaches based on a trainable parameter, cross-validation, and \ac{BLA} have been proposed to ensure the practical applicability of the normalization. Based on simulation studies, we have shown that the proposed methodologies enable to improve model estimation with \ac{NN}-based state-space modeling  methods.



\bibliography{root}

\end{document}